\documentclass{amsart}
\usepackage{amssymb, latexsym}

\newcommand{\beqa}{\begin{eqnarray*}}
\newcommand{\eeqa}{\end{eqnarray*}}
\newcommand{\beqn}{\begin{eqnarray}}
\newcommand{\eeqn}{\end{eqnarray}}

\newcommand{\iy}{\infty}

\newcommand{\Ha}{\mathbb H}

\newcommand{\al}{\alpha}

\newcommand{\e}{\varepsilon}

\newcommand{\Om}{ \Omega}

\newcounter{cnt1}
\newcounter{cnt2}
\newcounter{cnt3}
\newcommand{\blr}{\begin{list}{$($\roman{cnt1}$)$}
 {\usecounter{cnt1} \setlength{\topsep}{0pt}
 \setlength{\itemsep}{0pt}}}
\newcommand{\bla}{\begin{list}{$($\alph{cnt2}$)$}
 {\usecounter{cnt2} \setlength{\topsep}{0pt}
 \setlength{\itemsep}{0pt}}}
\newcommand{\bln}{\begin{list}{$($\arabic{cnt3}$)$}
 {\usecounter{cnt3} \setlength{\topsep}{0pt}
 \setlength{\itemsep}{0pt}}}
\newcommand{\el}{\end{list}}
\newtheorem{thm}{Theorem}
\newtheorem{lem}[thm]{Lemma}
\newtheorem{cor}[thm]{Corollary}
\newtheorem{ex}[thm]{Example}

\newtheorem{Def}[thm]{Definition}

\newtheorem{rem}[thm]{Remark}
\newcommand{\Rem}{\begin{rem} \rm}
\newcommand{\bdfn}{\begin{Def} \rm}
\newcommand{\edfn}{\end{Def}}

\newcommand{\ba}{\begin{array}}
\newcommand{\ea}{\end{array}}

\newtheorem*{acknowledgements}{Acknowledgements}
\date{}
\begin{document}
\title{\bf Global (in time) Solutions to The 3D-Navier-Stokes Equations on ${\mathbb{R}}^3$}
\author[Gill]{T. L. Gill}
\address[Tepper L. Gill]{ Department of Electrical Engineering, Howard University\\
Washington DC 20059 \\ USA, {\it E-mail~:} {\tt tgill@howard.edu}}
\author[Zachary]{W. W. Zachary}
\address[Woodford W. Zachary]{ Department of Electrical Engineering \\ Howard
University\\ Washington DC 20059 \\ USA, {\it E-mail~:} {\tt
wwzachary@earthlink.net}}
\date{}
\begin{abstract}  In two recent papers (\cite{GZ1} \cite{GZ2}), we provided solutions to the well-known unsolved problem of constructing sufficiency classes of functions in ${\mathbb H}{[{\mathbb {R}}^3]^3}$ and ${\mathbb V}{[{\mathbb {R}}^3]^3}$, which would allow global, in time, strong solutions to the three-dimensional Navier-Stokes equations. These equations describe the time evolution of the fluid velocity and pressure of an incompressible viscous homogeneous Newtonian fluid in terms of a given initial velocity and given external body forces.  In both previous papers, our solution was restricted to functions defined on a bounded open domain of class $\mathbb{C}^3 $ contained in ${\mathbb {R}}^3 $.  In this paper, we study this problem for functions defined on all of ${\mathbb {R}}^3 $. We prove that, under appropriate conditions, there exists a positive constant $a$ and a number $ {{\bf{u}}_ +}$, depending only on the domain, the viscosity, the body forces and the eigenvalues of the \textquotedblleft{Hermite}" Stokes operator (defined below) such that, for all functions in a dense set $\mathbb{D}$ contained in the closed ball ${{\mathbb B} ( {\mathbb {R}}^3 )}$ of radius $ (1/2){\bf{u}_ +} $ in ${\mathbb {H}}[ {\mathbb {R}}^3 ]^3$, the Navier-Stokes equations have unique strong solutions in ${\mathbb C}^{1} \left( {(0,\infty ),{\mathbb {H}}[ {\mathbb {R}}^3 ]^3} \right)$.
\end{abstract}
\maketitle
\section*{Introduction} Let ${{\mathbb L}^{{2}} [ {\mathbb {R}}^3 ]^3 }$ be the real Hilbert space of square integrable functions on ${\mathbb {R}}^3 $ with values in ${\mathbb {R}}^3$, and let ${\mathbb {H}}_0[ {\mathbb {R}}^3 ]^3$ be the completion of the set of functions in $\left\{ {{\bf{u}} \in \mathbb {C}_0^\infty  [ {\mathbb {R}}^3 ]^3 \left. {} \right|\,\nabla  \cdot {\bf{u}} = 0} \right\}$ which vanish at infinity with respect to the inner product of ${{\mathbb L}^2 [ {\mathbb {R}}^3 ]^3} $, and let ${\mathbb{V}}_0[ {\mathbb {R}}^3 ]^3$ be the completion of the above functions which vanish at infinity with respect to the inner product of $\mathbb{H}_0^1[ {\mathbb {R}}^3 ]$, the functions in ${\mathbb{H}}_0 [ {\mathbb {R}}^3 ]^3$ with weak derivatives in $(\mathbb{L}_{}^2 [ {\mathbb {R}}^3 ])^3 $.  The global in time classical Navier-Stokes initial-value problem (on $ \mathbb{R}^3 {\text{ and all }}T > 0$) is to find  functions ${\mathbf{u}}:[0,T] \times {\mathbb {R}}^3  \to \mathbb{R}^3$ and $p:[0,T] \times {\mathbb {R}}^3  \to \mathbb{R}$ such that
\beqn
\begin{gathered}
  \partial _t {\mathbf{u}} + ({\mathbf{u}} \cdot \nabla ){\mathbf{u}} - \nu \Delta {\mathbf{u}} + \nabla p = {\mathbf{f}}(t){\text{ in (}}0,T) \times {\mathbb {R}}^3 , \hfill \\
  {\text{                              }}\nabla  \cdot {\mathbf{u}} = 0{\text{ in (}}0,T) \times {\mathbb {R}}^3 {\text{ (in the weak sense),}} \hfill \\
  {\text{                              }}\mathop {\lim }\limits_{\left\| {\mathbf{x}} \right\| \to \infty } {\mathbf{u}}(t,{\mathbf{x}}) = 0{\text{ on }}\left( {0,T} \right) \times \mathbb{R}^3, \hfill \\
  {\text{                              }}{\mathbf{u}}(0,{\mathbf{x}}) = {\mathbf{u}}_0 ({\mathbf{x}}){\text{ in }}{\mathbb {R}}^3 . \hfill \\ 
\end{gathered} 
\eeqn
The equations describe the time evolution of the fluid velocity ${\mathbf{u}}({\mathbf{x}},t)$ and the pressure $p$ of an incompressible viscous homogeneous Newtonian fluid with constant viscosity coefficient $\nu $ in terms of a given initial velocity ${\mathbf{u}}_0 ({\mathbf{x}})$ and given external body forces
${\mathbf{f}}({\mathbf{x}},t)$.  (Note that our third condition, $\mathop {\lim }\limits_{\left\| {\mathbf{x}} \right\| \to \infty } {\mathbf{u}}(t,{\mathbf{x}}) = 0{\text{ on }}\left( {0,T} \right) \times \mathbb{R}^3$, is natural in this case since it is well-known that $\mathbb{H}_0^k [ {\mathbb {R}}^3 ]^3=\mathbb{H}^k [ {\mathbb {R}}^3 ]^3$ (see Stein \cite{S} or \cite{SY}.) 

\section*{Purpose}
Let $\mathbb{P}$ be the (Leray) orthogonal projection of 
$(\mathbb{L}^2 [ {\mathbb {R}}^3 ])^3$ 
onto ${{\mathbb{H}}_0}[ {\mathbb {R}}^3]^3$ and define the Stokes operator by:  $ {\bf{Au}} = : -\mathbb{P} \Delta {\bf{u}}$, 
for ${\bf{u}} \in D({\bf{A}}) \subset {\mathbb{H}}_0^{2}[ {\mathbb {R}}^3]^3$, the domain of ${\bf{A}}$.  Let $ {\bf{Bu}} = : 1/2{\mathbb{P}}(- \Delta + |{\bf {x}}|^2) {\bf{u}}$ for ${\bf{u}} \in D({\bf{B}})$. 
We call $\bf B$ the Hermite-Stokes operator. The purpose of this paper is to prove that there exists a number $ {{\bf{u}}_ +} $, depending only on ${\bf{A}}$, ${\bf{B}}$,  $f$,  $\nu $ and $ {\mathbb {R}}^3 $, such that, for all functions in 
$
\mathbb{D} = D({\bf{A}}) \cap \mathbb{B}({\mathbb {R}}^3 ),
$
 where ${{\mathbb B}( {\mathbb {R}}^3 )}$ is the closed ball of radius 
$ {\mathbf{u}}_ + $ in ${{\mathbb H}_0( {\mathbb {R}}^3 )^3}$, the Navier-Stokes equations have unique strong solutions in 
$
{\bf{u}}\in L_{\text{loc}}^\infty[[0,\infty); {\mathbb {V}}_0( {\mathbb {R}}^3)^3]
\cap \mathbb{C}^1[(0,\infty);{\mathbb {H}}_0( {\mathbb {R}}^3 )^3.$
\section*{Preliminaries}
In terms of notation and convention, we follow Sell and You \cite{SY}.   In order to simplify notation, we let ${\mathbb{H}}$  denote ${{\mathbb{H}}_0}[ {\mathbb {R}}^3]^3$ and ${\mathbb{V}}$ denote ${{\mathbb{V}}_0}[ {\mathbb {R}}^3]^3$. Our use of the Fourier transform follows the definition of Rudin \cite{RU}: $\mathfrak{F}(h) = \tfrac{1}{{ [{2\pi }]^{3/2} }}\int_{\mathbb{R}^3 } {e^{i{\mathbf{x}} \cdot {\mathbf{y}}} h({\mathbf{y}})d{\mathbf{y}}} $, so that no factors of ${2\pi } $ appear in the transform pairs. In order to simplify our proofs, we always assume that all functions $\bf u$, $\bf v$ are in $D(\bf A$) and, as in \cite{GZ2}, we take $c=max\{c_i\}$, where $c_i$ is one of the nine positive constants that appear on pages 363-367 in \cite{SY}.  It will also be convenient to use the fact that the norms of $\mathbb{V}$ and $\mathbb{V}^{ - 1}$ are equivalent in their respective  graph norms relative to $\mathbb{H}$.  
\section*{The Stokes Operator}
It is known that ${\bf{A}}$ is a nonnegative linear operator which generates an analytic contraction semigroup.  It follows that the fractional powers ${\bf{A}}^{1/2}$ and ${\bf{A}}^{ - 1/2}$ are well defined.  Moreover, it is also known (cf., \cite{SY}, \cite{T1}) that the norms $\left\| {{\bf{A}}^{1/2} {\bf{u}}} \right\|_{\mathbb{H}}$  and $\left\| {{\bf{A}}^{ - 1/2} {\bf{u}}} \right\|_{\mathbb{H}}$ are equivalent to the corresponding norms induced by the Sobolev space $(H^1 [ {\mathbb {R}}^3 ])^3 $, so that: 
\begin{equation}
\left\| {\mathbf{u}} \right\|_{\mathbb{V}}  \equiv \left\| {{\mathbf{A}}^{1/2} {\mathbf{u}}} \right\|_{\mathbb{H}} {\text{ and  }}\left\| {\mathbf{u}} \right\|_{\mathbb{V}^{-1} }  \equiv \left\| {{\mathbf{A}}^{ - 1/2} {\mathbf{u}}} \right\|_{\mathbb{H}} .
\end{equation}
In addition,  $\mathbf{A}$ is an isomorphism from $
D({\mathbf{A}})\xrightarrow{{onto}}D({\mathbf{A}}^{ - 1})$.   Furthermore, the embeddings $\mathbb{V} \to \mathbb{H} \to \mathbb{V}^{ - 1} $
 are continuous, and it is easy to see that  $\mathbf{A}^{ - 1} $ is the projection of an operator represented by the Riesz potential, mapping $D({\mathbf{A}}^{ - 1})$ onto $D({\mathbf{A}})$ (see Stein \cite{S}). Applying the Leray projection to equation (1), with 
${\mathbf{C}}({\mathbf{u}},{\mathbf{u}}) = \mathbb{P}({\mathbf{u}} \cdot \nabla ){\mathbf{u}}$, we can recast equation (1) in the standard form:
\beqn
\begin{gathered}
  \partial _t {\mathbf{u}} =  - \nu {\mathbf{Au}} - {\mathbf{C}}({\mathbf{u}},{\mathbf{u}}) + \mathbb{P}{\mathbf{f}}(t){\text{ in (}}0,T) \times {\mathbb {R}}^3 , \hfill \\
 {\text{                              }}\nabla  \cdot {\mathbf{u}} = 0{\text{ in (}}0,T) \times {\mathbb {R}}^3 , \hfill \\
  {\text{                              }}\mathop {\lim }\limits_{\left\| {\mathbf{x}} \right\| \to \infty } {\mathbf{u}}(t,{\mathbf{x}}) = 0{\text{ on }}\left( {0,T} \right) \times \mathbb{R}^3, \hfill \\
  {\text{                              }}{\mathbf{u}}(0,{\mathbf{x}}) = {\mathbf{u}}_0 ({\mathbf{x}}){\text{ in }}{\mathbb {R}}^3, \hfill \\ 
\end{gathered} 
\eeqn
where we have used the fact that the orthogonal complement of ${\Ha}[{\mathbb {R}}^3 ] $ relative to $(\mathbb{L}^{2}) [{\mathbb {R}}^3 ])^3 $ is $\{ {\mathbf{v}}\,:\;{\mathbf{v}} = \nabla q,\;q \in (H^1 [{\mathbb {R}}^3 ])^3 \} $ to eliminate the pressure term (see Galdi [GA] or [SY, T1, T2]). 
Theorem 1 below will be used to get our basic estimate in Theorem 3.  This result is a simple extension of the bounded domain case first proved by  Constantin and Foi\c{a}s \cite{CF}.
\begin{thm}   Let ${\alpha_i,1 \le i \le 3}$, satisfy 
$ {0 \le \alpha_1\le 3}$, ${0 \le \alpha_2 \le 2}$, ${0 \le \alpha_3 \leq 3}$, with ${ \alpha_1+\alpha_2+\alpha_3 \ge 3/2}$ and
\beqa
(\alpha _1 ,\alpha _2 ,\alpha _3 ) \notin \left\{ {(3/2,0,0),(0,3/2,0),(0,0,3/2)} \right\}.
\eeqa
Then there is a positive constant $c=c(\al_i)$ such that
$$
\left| {\left\langle {{\mathbf{C}}({\mathbf{u}},{\mathbf{v}}),{\mathbf{w}}} \right\rangle _\mathbb{H} } \right| \leqslant c \left\| {{\mathbf{A}}^{\alpha _1 /2} {\mathbf{u}}} \right\|_\mathbb{H} \left\| {{\mathbf{A}}^{(1 + \alpha _2 )/2} {\mathbf{v}}} \right\|_\mathbb{H} \left\| {{\mathbf{A}}^{\alpha _3 /2} {\mathbf{w}}} \right\|_\mathbb{H}. 
$$
\end{thm}
We shall make use of the following interpolation inequality:  (see Sell and You \cite{SY}, page 363) 
$$
\left\| {{\mathbf{A}}^\gamma  {\mathbf{u}}} \right\|_\mathbb{H}  \leqslant c \left\| {{\mathbf{A}}^\alpha  {\mathbf{u}}} \right\|_\mathbb{H}^\theta  \left\| {{\mathbf{A}}^\beta  {\mathbf{u}}} \right\|_\mathbb{H}^{(1 - \theta )} 
$$
for all ${\mathbf{u}} \in D({\mathbf{A}}^\alpha  )$, where $
\gamma  = \theta \alpha  + (1 - \theta )\beta ,{\text{  }}\alpha, \beta ,\gamma  \in \mathbb{R}$, $0 \le \theta \le 1$ and $\beta  \leqslant \alpha $. 

\section*{The Hermite-Stokes Operator}
The operator ${\bf {\hat B}}=1/2(-\Delta + |\bf x|^2)$ is the three-dimensional version of the standard harmonic oscillator operator, which generates the Hermite functions (products of the Hermite polynomials by $e^{-x^{2}/2}$) as eigenfunctions for the
eigenvalue problem on $\mathbb {R}$, ( see Hermite \cite{HR}, Appell and Kam\'{e} de F\'{e}riet \cite{AK}, and Magnus, Oberhettinger and Soni \cite{MOS}).  It is easy to show directly, by separation of variables, that the solution to the 3-dimensional problem is the product of the solutions to the 1-dimensional problem, while the eigenvalues for the 3-dimensional Hermite polynomials are the sums of those for the 1-dimensional polynomials.  Furthermore, ${\bf {\hat B}}$, and hence ${\mathbf{B}} = \mathbb{P}{\mathbf{\hat B}}$, is positive with a compact inverse, while ${\bf {A}}$ has an unbounded inverse on $\mathbb {H}_0(\mathbb {R}^3)^3$.   It turns out that $\bf{\hat B}$ is \textquotedblleft{natural}" for  ${\mathbb{R}}^3$ in the sense that it is the only positive self-adjoint (sectorial) operator of lowest degree that is invariant under both rotations and Fourier transformations. (This is actually true for ${\mathbb{R}}^n,\; n \ge 1$.)

We will have need of the fact that every function $
{\text{ }}{\mathbf{h}}(t) \in \mathbb{H}$ has an expansion in terms of the eigenfunctions of ${\mathbf{B}}$ so that, for example,  $
{\mathbf{B}}^{ -\beta} {\text{ }}{\mathbf{h}}(t) = \sum\nolimits_{k = 1}^\infty  {\lambda _k^{ -\beta} h_k (t){\mathbf{e}}^k ({\mathbf{x}})}$ and, from here, it is easy to see that $
\left\| {{\mathbf{B}}^{ -\beta} {\text{ }}{\mathbf{h}}(t)} \right\|_{\mathbb{H}}  \leq \lambda _1^{ -\beta} \left\| {{\text{ }}{\mathbf{h}}(t)} \right\|_{\mathbb{H}} $, where ${\lambda} _1^{-1}$ is the largest eigenvalue of ${\mathbf{B}}^{-1}$.  We also need the following result for  our basic Theorem. 

\begin{lem} $D(\bf A)=D(\bf B)$.
\begin{proof}
If we define a norm on $D({\mathbf{A}})$
 by $\left\| {\mathbf{u}} \right\|_{\mathbf{A}}  = \left\| {{\mathbf{Au}}} \right\|_\mathbb{H} $, then $\left( {D({\mathbf{A}})\,,\,\,\left\| {\, \cdot \,} \right\|_{\mathbf{A}} } \right)$ is a Hilbert space.   Now note that the Fourier transform $\mathfrak{F}( \cdot )$ is an isometric isomorphism on $\left( {D({\mathbf{A}})\,,\,\,\left\| {\, \cdot \,} \right\|_{\mathbf{A}} } \right)$ to $\left( {D(\mathbb{P}\left| {\mathbf{x}} \right|^2 )\,,\,\,\left\| {\, \cdot \,} \right\|_{\mathbf{A}} } \right),$ since $\left\| {{\mathbf{Au}}} \right\|_\mathbb{H}  = \left\| {\mathfrak{F}({\mathbf{Au}})} \right\|_\mathbb{H}  = \left\| {\mathbb{P}\left| {\mathbf{x}} \right|^2 {\mathbf{\hat u}}} \right\|_\mathbb{H} $.  It is now easy to see that $D({\mathbf{A}}) = D(\mathbb{P}\left| {\mathbf{x}} \right|^2 )$.  From this, it follows that $D({\mathbf{A}}) = D({\mathbf{B}})$.
\end{proof}
\end{lem}

It follows from the above lemma that $({{\mathbf{AB}})^{ - \delta}}$ is bounded for $\delta >0$.  The following estimate is equation 61.24.1 on page 366 in  Sell and You \cite{SY}.  If we set $\alpha_1=1, \alpha_2=1/2$, and  $\alpha_3=0$ in Theorem 1, along with the interpolation inequality, we get that
\beqn
\left| {\left\langle {{\mathbf{C}}({\mathbf{u}},{\mathbf{v}}),{\mathbf{w}}} \right\rangle _\mathbb{H} } \right| \leqslant c \left\| {{\mathbf{A}}^{1/2} {\mathbf{u}}} \right\|_\mathbb{H} \left\| {{\mathbf{A}} {\mathbf{v}}} \right\|_\mathbb{H} \left\| {\mathbf{w}} \right\|_\mathbb{H}. 
\eeqn
\begin{thm} Let $\mathbf{u,v,w} \in \mathbb{H}$, and let $\e >0$ be arbitrary.  Then, for $\delta =1/4+\e/2$, we have that:
\beqn
\left| {\left\langle ({{\mathbf{AB}})^{ - (1 + \delta )} {\mathbf{C}}({\mathbf{u}},{\mathbf{v}}),{\mathbf{w}}} \right\rangle _\mathbb{H} } \right| \leqslant {c{\lambda _1^{ - (1 + \delta )}}  } \left\| {\mathbf{u}} \right\|_\mathbb{H} \left\| {\mathbf{v}} \right\|_\mathbb{H} \left\| {\mathbf{w}} \right\|_\mathbb{H}. 
\eeqn
\end{thm}
\begin{proof}
Using the self-adjoint property of $\mathbf{A}$, and integration by parts, we have
$$
\left\langle {{\mathbf{A}}^{ - \beta } {\mathbf{C}}({\mathbf{u}},{\mathbf{v}}),{\mathbf{h}}} \right\rangle _\mathbb{H}  = \left\langle {{\mathbf{C}}({\mathbf{u}},{\mathbf{v}}),{\mathbf{A}}^{ - \beta } {\mathbf{h}}} \right\rangle _\mathbb{H}  =  - \left\langle {{\mathbf{C}}({\mathbf{u}},{\mathbf{A}}^{ - \beta } {\mathbf{h}}),{\mathbf{v}}} \right\rangle _\mathbb{H}.
$$
It now follows from Theorem 1 that:
$$
\left| {\left\langle {{\mathbf{A}}^{ - \beta } {\mathbf{C}}({\mathbf{u}},{\mathbf{v}}),{\mathbf{h}}} \right\rangle _\mathbb{H} } \right| \leqslant c \left\| {{\mathbf{A}}^{\alpha _1 /2} {\mathbf{u}}} \right\|_\mathbb{H} \left\| {{\mathbf{A}}^{ - \beta  + (1 + \alpha _2 )/2} {\mathbf{h}}} \right\|_\mathbb{H} \left\| {{\mathbf{A}}^{\alpha _3 /2} {\mathbf{v}}} \right\|_\mathbb{H}. 
$$
If we set $\beta  = 1 + \delta ,{\text{ }}\alpha _1 {\text{ = }}\alpha _3 {\text{ = 0,}}$ we have 
$$\left| {\left\langle {{\mathbf{A}}^{ - (1 + \delta )} {\mathbf{C}}({\mathbf{u}},{\mathbf{v}}),{\mathbf{h}}} \right\rangle _\mathbb{H} } \right| \leqslant c\left\| {\mathbf{u}} \right\|_\mathbb{H} \left\| {\mathbf{v}} \right\|_\mathbb{H} \left\| {{\mathbf{A}}^{(\alpha _2  - 1 - 2\delta )/2} {\mathbf{h}}} \right\|_\mathbb{H} .
$$
With $\delta  = 1/4 + \varepsilon /2,$ we get that, for the last term to reduce to $\left\| {\mathbf{h}} \right\|_\mathbb{H} $, we can set $\alpha _2  = 3/2 + \varepsilon $.  It follows that the conditions of Theorem 1 are satisfied if $3/2 + \varepsilon  < 2$.  Thus, it suffices to assume that $\varepsilon <1/2$, which we will do in the rest of the paper without comment.  Our proof is completed by taking ${\mathbf{h}}={\mathbf{B}}^{-\beta}{\mathbf{w}}$, and the fact that $\left\| {\mathbf{B}}^{-\beta}{\mathbf{w}} \right\|_\mathbb{H}\leq {\lambda _1^{-\beta}}\left\| {\mathbf{w}} \right\|_\mathbb{H}$. 
\end{proof}
\begin{ex}  If we use Theorem 1, with $\alpha_1=5/4$, $\alpha_2=1/4,$ and $\alpha_3=0 $, along with the interpolation inequality, and the fact that $\left\| {{\mathbf{A}}^{1/2} {\mathbf{u}}} \right\|_\mathbb{H}  \leqslant \left\| {{\mathbf{Au}}} \right\|_\mathbb{H}$ we have that, for all $
{\mathbf{u}},{\mathbf{v}} \in D({\mathbf{A}})$,
\beqn
\begin{gathered}
  \left\| {{\mathbf{C}}({\mathbf{u}},{\mathbf{v}})} \right\|_\mathbb{H}  \leqslant c \left\| {{\mathbf{A}}^{1/2} {\mathbf{u}}} \right\|_\mathbb{H}^{3/4} \left\| {{\mathbf{Au}}} \right\|_\mathbb{H}^{1/4} \left\| {{\mathbf{A}}^{1/2} {\mathbf{v}}} \right\|_\mathbb{H}^{3/4} \left\| {{\mathbf{Av}}} \right\|_\mathbb{H}^{1/4}  \hfill \\
  {\text{                }} \leqslant c \left\| {{\mathbf{Au}}} \right\|_\mathbb{H} \left\| {{\mathbf{Av}}} \right\|_\mathbb{H} . \hfill \\ 
\end{gathered} 
\eeqn
A better estimate is possible, but for our use, equation (6) will suffice.
\end{ex} 

\begin{Def}  We say that the operator ${\mathbf{J}}( \cdot ,t)$ is (for each $t$) 
\begin{enumerate}
\item
0-Dissipative if $
\left\langle {{\mathbf{J}}({\mathbf{u}},t),{\mathbf{u}}} \right\rangle _{\mathbb{H}}  \le 0$.
\item
Dissipative if 
$\left\langle {{\mathbf{J}}({\mathbf{u}},t) - {\mathbf{J}}({\mathbf{v}},t),{\mathbf{u}} - {\mathbf{v}}} \right\rangle _{\mathbb{H}}  \le 0$.
\item
Strongly dissipative if there  exists an $ \alpha > 0$ such that
$$
\left\langle {{\mathbf{J}}({\mathbf{u}},t) - {\mathbf{J}}({\mathbf{v}},t),{\mathbf{u}} - {\mathbf{v}}} \right\rangle _{\mathbb{H}}  \le  - \alpha \left\| {{\mathbf{u}} - {\mathbf{v}}} \right\|_{\mathbb{H}}^2. 
$$
\item
Uniformly dissipative if there exists a strictly monotone increasing function $a(t)$
 with $a(0) = 0$, $\lim _{t \to \infty } a(t) = \infty $, and: 
$$
\left\langle {{\mathbf{J}}({\mathbf{u}},t) - {\mathbf{J}}({\mathbf{v}},t),{\mathbf{u}} - {\mathbf{v}}} \right\rangle _{\mathbb{H}}  \le - a\left( {\left\| {{\mathbf{u}} - {\mathbf{v}}} \right\|_{\mathbb{H}} } \right)\left\| {{\mathbf{u}} - {\mathbf{v}}} \right\|_{\mathbb{H}} .
$$
\end{enumerate}
\end{Def}
Note that, if ${\mathbf{J}}( \cdot ,t)$ is a linear operator, definitions 1) and 2) coincide.  Theorem 6 below is essentially due to Browder \cite{B},  see Zeidler {\cite[Corollary 32.27, page 868 and Corollary 32.35, page 887 in, Vol. IIB]{Z}}, while Theorem 7 is from Miyadera \cite[p. 185, Theorem 6.20]{M}, and is a modification of the Crandall-Liggett Theorem \cite{CL} (see the appendix to the first section of \cite{CL}) . 
\begin{thm} Let $\mathbb{B}[ {\mathbb R}^3 ]$ be a closed, bounded, convex subset of $
\mathbb{H}[  {\mathbb R}^3 ]$.  If ${\mathbf{J}}( \cdot ,t):\mathbb{B}[  {\mathbb R}^3 ] \to \mathbb{H}[  {\mathbb R}^3 ]$ is closed and strongly dissipative for each fixed $t \ge 0$ then, for each ${\mathbf{b}} \in \mathbb{B}[  {\mathbb R}^3 ]$, there is a 
${\mathbf{u}} \in \mathbb{B}[  {\mathbb R}^3 ]$ with ${\mathbf{J}}({\mathbf{u}},t) = {\mathbf{b}}$ (e.g., the range, $
Ran{\text{[}}{\mathbf{J}}( \cdot ,t)] \supset \mathbb{B}[  {\mathbb R}^3 ]$).
\end{thm}	
\begin{thm} Let ${\text{\{ }}\mathcal{A}(t), t \in I = [0,\infty ){\text{\} }}$ be a family of operators defined on $\mathbb{H}[  {\mathbb R}^3 ]$ with domains $
D(\mathcal{A}(t)) = D$, independent of $t$. We assume that $\mathbb{D} = D \cap \mathbb{B}[  {\mathbb R}^3 ]$ is a closed convex set (in an appropriate topology):
\begin{enumerate}
\item
The operator $\mathcal{A}(t)$ is the generator of a contraction semigroup for each
$t \in I$.
\item
The function $\mathcal{A}(t){\mathbf{u}}$ is continuous in both variables on $
I \times \mathbb{D}$.
\end{enumerate}
Then, for every ${\mathbf{u}}_0  \in \mathbb{D}$, the problem 
$\partial _t {\mathbf{u}}(t,{\mathbf{x}}) = \mathcal{A}(t){\mathbf{u}}(t,{\mathbf{x}})$, 
${\mathbf{u}}(0,{\mathbf{x}}) = {\mathbf{u}}_0 ({\mathbf{x}})$, has a unique solution 
${\mathbf{u}}(t,{\mathbf{x}}) \in \mathbb{C}^1 (I;\mathbb{D})$.
\end{thm} 
\section*{M-Dissipative Conditions} 
Let us assume that 
$
{\mathbf{f}}(t) \in L^\infty[[0,\infty); {\mathbb H}]
$
and is Lipschitz continuous in $t$, with $\left\| {{\mathbf{f}}(t) - {\mathbf{f}}(\tau )} \right\|_{\mathbb{H}}  \le d\left| {t - \tau } \right|^\theta,{\text{ }}d > 0,{\text{ }}0 < \theta  < 1$.  With $\delta$ as in Theorem 3, we can rewrite equation (3) in the form:
\beqn
\begin{gathered}
  \partial _t {\mathbf{u}} = \nu ({\bf{AB}})^{1 + \delta } {\mathbf{J}}({\mathbf{u}},t){\text{ in (}}0,T) \times \Omega , \hfill \\
  {\mathbf{J}}({\mathbf{u}},t) =  - {\bf B}^{-(1+ \delta)}{\mathbf{A}}^{ - \delta } {\mathbf{u}} - \nu ^{ - 1}  ({\bf AB})^{-(1+ \delta)} {\mathbf{C}}({\mathbf{u}},{\mathbf{u}}) + \nu ^{ - 1}   ({\bf AB})^{-(1+ \delta)} \mathbb{P}{\mathbf{f}}(t). \hfill \\ 
\end{gathered} 
\eeqn
\section*{Approach} 	We begin with a study of the operator ${\mathbf{J}}( \cdot ,t)$, for fixed $t$, and seek conditions depending on ${\mathbf{A}}, {\bf B}, {\text{ }}\nu ,{\text{ }}  {\text{ and }}{\mathbf{f}}(t)$ which guarantee that ${\mathbf{J}}( \cdot ,t)$ is m-dissipative for each $t$.  Clearly $
{\mathbf{J}}( \cdot ,t)\!: D[({\bf{AB})^{(1+\delta)}}] \xrightarrow{{onto}} D[{(\mathbf{AB})^{(1+ \delta)}}]$ and, since $ \nu (\mathbf{AB})^{(1+ \delta)} $ is a closed positive (m-accretive) operator (so that $ - {(\mathbf{AB})^{(1+ \delta)}}$ generates a linear contraction semigroup), we expect that $\nu {(\mathbf{AB})}^{(1+ \delta)}J( \cdot ,t)$ will be m-dissipative for each $t$.  
\begin{thm} For $t \in I=[0, \infty)$ and, for each fixed ${\mathbf{u}} \in \mathbb{H}$, $
{\mathbf{J}}({\mathbf{u}},t)$ is Lipschitz continuous, with $
\left\| {{\mathbf{J}}({\mathbf{u}},t) - {\mathbf{J}}({\mathbf{u}},\tau )} \right\|_{\mathbb{H}}  \le d'\left| {t - \tau } \right|^\theta$, where $d' = d{{\nu}^{-1}}a^{ -(1+ \delta)} $, $d$ is the Lipschitz constant for the function ${\mathbf{f}}(t)$ and $a^{ -(1+ \delta)}=\left\|{(\mathbf{AB})}^{-(1+ \delta)}\right\|_{\mathbb{H}}$. 
\end{thm}
\begin{proof}
For fixed ${\mathbf{u}} \in \mathbb{H}$, 
\[
\begin{gathered}
  \left\| {{\mathbf{J}}({\mathbf{u}},t) - {\mathbf{J}}({\mathbf{u}},\tau )} \right\|_{\mathbb{H}}  = \nu ^{ - 1} \left\| {({\mathbf{AB}})^{ -(1+ \delta)} {\text{[}}\mathbb{P}{\mathbf{f}}(t) - \mathbb{P}{\mathbf{f}}(\tau )]} \right\|_{\mathbb{H}}  \hfill \\
  {\text{                                  }} \leq d{\nu^{-1}}a^{ -(1+ \delta)} \left| {t - \tau } \right|^\theta   = d'\left| {t - \tau } \right|^\theta . \hfill \\ 
\end{gathered} 
\]
\end{proof}
\section*{Main Results} 
\begin{thm} Let $f = \sup _{t \in {\mathbf{R}}^ +  } \left\| {\mathbb{P}{\mathbf{f}}(t)} \right\|_{\mathbb{H}}  < \infty $, then there exists a positive constant ${\mathbf{u}}_ +  $, depending only on $f$, ${\mathbf{A}}$, ${\mathbf{B}}$ and $\nu $  such that, for all ${\mathbf{u}}$ with $
\left\| {\mathbf{u}} \right\|_{\mathbb{H}}  \le {\mathbf{u}}_ +  $, ${\mathbf{J}}( \cdot ,t)$ is strongly dissipative. \end{thm}
\begin{proof} The proof of our first assertion has two parts. First, we require that the nonlinear operator ${\mathbf{J}}( \cdot ,t)$
 be 0-dissipative, which gives us an upper bound ${\mathbf{u}}_ +  $ in terms of the norm (e.g., $\left\| {\mathbf{u}} \right\|_{\mathbb{H}}  \leqslant {\mathbf{u}}_ + $ ).  We then use this part, and the fact that $\left\| {\mathbf{u}} \right\|_{\mathbb{H}}  \leqslant \left\| {\mathbf{Au}} \right\|_{\mathbb{H}} $, to show that ${\mathbf{J}}( \cdot ,t)$ is strongly dissipative on the closed ball, $ \mathbb{B}{_+} = \left\{ {{\mathbf{u}} \in \mathbb{H}:\left\| {\mathbf{Au}} \right\|_{\mathbb{H}}  \leqslant (1/2){\mathbf{u}}_ +  } \right\}$.  

Part 1) 
From equation (5), we consider the expression
\beqa
\begin{gathered}
  \left\langle {{\mathbf{J}}({\mathbf{u}},t),({\mathbf{AB}})^{ - \delta }{\mathbf{u}}} \right\rangle _{\mathbb{H}}  =  - \left\langle {{\mathbf{B}}^{ - 1}({\mathbf{AB}})^{ - \delta } {\mathbf{u}},({\mathbf{AB}})^{ - \delta }{\mathbf{u}}} \right\rangle _\mathbb{H}  \hfill \\
  \;\;\;\;\;\;\;\;\;\;\;\;\;\;\;\;\;\;\;\;\;\;\;\;\;\;\;\;\;\;\;\;{\text{                    }} + \nu ^{ - 1} \left\langle { - ({\mathbf{AB}})^{ - (1+\delta) } {\mathbf{C}}({\mathbf{u}},{\mathbf{u}}) + ({\mathbf{AB}})^{ - (1+\delta) } \mathbb{P}{\mathbf{f}}(t),({\mathbf{AB}})^{ - \delta }{\mathbf{u}}} \right\rangle _\mathbb{H}  \hfill \\
  {\text{                    }} =  - \left\| {{\bf B}^{-1/2}({\mathbf{AB}})^{ - \delta} {\mathbf{u}}} \right\|_\mathbb{H}^2  - \nu ^{ - 1} \left\langle {({\mathbf{AB}})^{ - (1+\delta) } {\mathbf{C}}({\mathbf{u}},{\mathbf{u}}),({\mathbf{AB}})^{ - \delta }{\mathbf{u}}} \right\rangle _\mathbb{H}  + \nu ^{ - 1} \left\langle {({\mathbf{AB}})^{ - (1+\delta) } \mathbb{P}{\mathbf{f}}(t),({\mathbf{AB}})^{ - \delta }{\mathbf{u}}} \right\rangle _\mathbb{H} \hfill \\
  {\text{                    }} =  - \left\| {{\bf{B}}^{-1/2}({\mathbf{AB}})^{ - \delta } {\mathbf{u}}} \right\|_\mathbb{H}^2  - \nu ^{ - 1} \left\langle {{\mathbf{C}}(({\mathbf{AB}})^{ - (1+\delta) }{\mathbf{u}}, {\mathbf{u}}),({\mathbf{AB}})^{ - \delta }{\mathbf{u}}} \right\rangle _\mathbb{H}  + \nu ^{ - 1} \left\langle ({\mathbf{AB}})^{ - (1+\delta) }{\mathbb{P}{\mathbf{f}}(t), ({\mathbf{AB}})^{ - \delta }{\mathbf{u}}} \right\rangle _\mathbb{H}. \hfill \\ 
\end{gathered} 
\eeqa
It follows that
\beqa
\begin{gathered}
  \left\langle {{\mathbf{J}}({\mathbf{u}},t),({\mathbf{AB}})^{ - \delta }{\mathbf{u}}} \right\rangle _{\mathbb{H}}  \leqslant  - \left\|
   {\bf{B}}^{-1/2}({\mathbf{AB}})^{ - \delta} {\mathbf{u}} \right\|_\mathbb{H}^2  + \nu ^{ - 1} \left| {\left\langle {{\mathbf{C}}(({\mathbf{AB}})^{ - (1+\delta) }{\mathbf{u}}, {\mathbf{u}}),({\mathbf{AB}})^{ - \delta }{\mathbf{u}}} \right\rangle _\mathbb{H} } \right|  \hfill \\
  \;\;\;\;\;\;\;\;\;\;\;\;\;\;\;\;\;\;\;\;\;\;\;\;\;\;\;\;\;\;\;\;{\text{                     }}+ \nu ^{ - 1} a^{-(1+\delta)}f \left\| {({\mathbf{AB}})^{ - \delta } {\mathbf{u}}} \right\|_\mathbb{H}  \hfill \\
  {\text{                     }} \leqslant  - \left\| {\bf{B}}^{-1/2}({\mathbf{AB}})^{ - \delta} {\mathbf{u}} \right\|_\mathbb{H}^2  + ca^{-\delta}(\nu \lambda _1^{(1 + \delta )} )^{ - 1}\left\| {\mathbf{u}} \right\|_\mathbb{H}^3  + \nu ^{ - 1}a^{ - (1 + 2\delta )} f\left\| { {\mathbf{u}}} \right\|_\mathbb{H} . \hfill \\ 
\end{gathered} 
\eeqa 
In the last line, we used our estimate from Theorem 3.  We now choose the first eigenvalue $\lambda _n,\; n \ge 1$, and number ${\text{ }}\omega $ such that 
\begin{enumerate}
\item
$\lambda _n^{-1/2} a^{ - \delta } \left\| {\mathbf{u}} \right\|_\mathbb{H}  \leqslant \left\| {\bf{B}}^{-1/2}({\mathbf{AB}})^{ - \delta } {\mathbf{u}} \right\|_\mathbb{H}  \leqslant \lambda _1^{-1/2} a^{ - \delta } \left\| {\mathbf{u}} \right\|_\mathbb{H} ,$
\item
$\lambda _1^{ - \omega/2} a^{-\delta }\left\| {\mathbf{u}} \right\|_\mathbb{H}  \leqslant \left\| {\bf{B}}^{-1/2}({\mathbf{AB}})^{ - \delta } {\mathbf{u}} \right\|_\mathbb{H}  \leqslant \lambda _1^{ - 1/2} a^{-\delta}\left\| {\mathbf{u}} \right\|_\mathbb{H} ,$
\end{enumerate}
and let $\lambda _0^{ -1}  = \max \{ \lambda _n^{ - 1} ,\lambda _1^{ - \omega} \} $.  It then follows that $ - \lambda _0^{ - 1} a^{ - 2\delta}\left\| {\mathbf{u}} \right\|_\mathbb{H}^2  \geqslant  - \left\| {\bf{B}}^{-1/2}({\mathbf{AB}})^{ - \delta} {\mathbf{u}} \right\|_\mathbb{H}^2 $.  Thus, ${\mathbf{J}}( \cdot ,t)$
will be 0-dissipative if 
\beqa
- \lambda _0^{ - 1} a^{ - 2\delta} \left\| {\mathbf{u}} \right\|_\mathbb{H}^2  +  ca^{ - \delta}(\nu \lambda _1^{(1 + \delta )} )^{ - 1}\left\| {\mathbf{u}} \right\|_\mathbb{H}^3  + (\nu a^{(1 + 2\delta )} )^{ - 1} f\left\| {\mathbf{u}} \right\|_{\mathbb{H}}  \leqslant 0,
\eeqa
so that 
\beqn 
a^{-\delta}\left\| {\mathbf{u}} \right\|_{\mathbb{H}} \left[c(\nu \lambda _1^{(1 + \delta )} )^{ - 1}
\left\| {\mathbf{u}} \right\|_\mathbb{H}^2  - \lambda _0^{ - 1} a^{ - \delta}  
\left\| {\mathbf{u}} \right\|_\mathbb{H}  + (\nu a^{(1 + \delta )} )^{ - 1} f \right] \leqslant 0.
\eeqn
Since $\left\| {\mathbf{u}} \right\|_\mathbb{H}  > 0$, we have that ${\mathbf{J}}( \cdot ,t)$ is 0-dissipative if
\beqa
c(\nu \lambda _1^{(1 + \delta )} )^{ - 1}
\left\| {\mathbf{u}} \right\|_\mathbb{H}^2  - \lambda _0^{ - 1} a^{ - \delta}  
\left\| {\mathbf{u}} \right\|_\mathbb{H}  + (\nu a^{(1 + \delta )} )^{ - 1} f  \leqslant 0.
\eeqa
Solving, we get that
\beqa
{\mathbf{u}}_ \pm   = \tfrac{\nu \lambda _1^{1+\delta}}
{2c \lambda _0 a^\delta  } \left\{ {1 \pm \sqrt {1 - ({{4c\lambda _0 ^{2} f)} \mathord{\left/
 {\vphantom {{4c\lambda _0 ^{2} f)} {(\nu ^2 a^{(1-\delta)} \lambda _1^{(1 + \delta )} )}}} \right.
 \kern-\nulldelimiterspace} {(\nu ^2  a^{(1-\delta)}\lambda _1^{(1 + \delta )} )}}} } \right\} = \tfrac{\nu \lambda _1^{1+\delta}}
{2c \lambda _0 a^\delta  } \left\{ {1 \pm \sqrt {1 - \gamma } } \right\},
\eeqa
where $
\gamma  = ({{4c\lambda _0 ^{2 } f)} \mathord{\left/
 {\vphantom {{4c\lambda _0 ^{2 } f)} {(\nu ^2 a^{(1-\delta)} \lambda _1^{(1 + \delta )} )}}} \right.
 \kern-\nulldelimiterspace} {(\nu ^2 a^{(1-\delta)} \lambda _1^{(1 + \delta )} )}}.
$
Since we want real distinct solutions, we must require that 
\beqa
\gamma  = ({{4c\lambda _0 ^{2 } f)} \mathord{\left/
 {\vphantom {{4c\lambda _0 ^{2 } f)} {(\nu ^2 a^{(1-\delta)} \lambda _1^{(1 + \delta )} )}}} \right.
 \kern-\nulldelimiterspace} {(\nu ^2 a^{(1-\delta)} \lambda _1^{(1 + \delta )} )}} < 1 \Rightarrow \nu ^2 a^{(1-\delta)}\lambda _1^{(1 + \delta )}  > 4c\lambda _0 ^{2 } f{\text{ }}  \hfill \\
  {\text{                    }} \Rightarrow {\text{  }}\nu  > 2\lambda _0  a^{-(1-\delta)/2} \lambda _1^{ - (1 + \delta )/2} (cf)^{1/2}. 
\eeqa
It follows that, if $\mathbb{P}{\mathbf{f}} \ne {\mathbf{0}}$, then 
$
{\mathbf{u}}_ -   < {\mathbf{u}}_ + $ , and our requirement that ${\mathbf{J}}$
 is 0-dissipative implies that, since our solution factors as 
$
(\left\| {\mathbf{u}} \right\|_{\mathbb{H}}  - {\mathbf{u}}_ +  )(\left\| {\mathbf{u}} \right\|_{\mathbb{H}}  - {\mathbf{u}}_ -  ) \le 0,
$
we must have that:
\beqa
\left\| {\mathbf{u}} \right\|_{\mathbb{H}}  - {\mathbf{u}}_ +   \le 0,{\text{  }}\left\| {\mathbf{u}} \right\|_{\mathbb{H}}  - {\mathbf{u}}_ -   \ge 0.
\eeqa
First observe that terms of the form $({\mathbf{AB}})^{ - \delta }{\mathbf{u}}$ are dense.  Then note that ${\mathbf{J}}({\mathbf{u}},t)$ is closed, and  the dissipative nature  of an operator is determined on a dense set.  It follows that, for  
${\text{ }}{\mathbf{u}}_ -   \le \left\| {\mathbf{u}} \right\|_{\mathbb{H}}  \le {\mathbf{u}}_ +  $, 
$
\left\langle {{\mathbf{J}}({\mathbf{u}},t),{\mathbf{u}}} \right\rangle _{\mathbb{H}}  \le 0$.  (It is clear that, when $
\mathbb{P}{\mathbf{f}}(t) = {\mathbf{0}}, {\mathbf{u}}_ -   = {\mathbf{0}}$, and ${\mathbf{u}}_ +   = \nu (c\lambda _0 a^\delta  )^{ - 1}{\lambda _1^{(1+\delta)}} $.)

Part 2): Now, for any ${\mathbf{u}},{\mathbf{v}} \in \mathbb{H}$  with $\max ({\text{ }}\left\| {\mathbf{Au}} \right\|_{\mathbb{H}} ,\left\| {\mathbf{Av}} \right\|_{\mathbb{H}} ) \le (1/2){{\mathbf{u}}_ +}  $, we have that   
\beqa
\begin{gathered}
  \left\langle {{\mathbf{J}}({\mathbf{u}},t) - {\mathbf{J}}({\mathbf{v}},t),({\mathbf{AB}})^{ - \delta }({\mathbf{u}} - {\mathbf{v}})} \right\rangle _{\mathbb{H}}  =  - \left\| {\bf{B}^{-1/2}({\mathbf{AB}})^{ - \delta } ({\mathbf{u}} - {\mathbf{v}})} \right\|_\mathbb{H}^2  \hfill \\
  {\text{                                                    }} - {\nu ^{ - 1}} \left\langle {({\mathbf{AB}})^{ - (1 + \delta )} [{\mathbf{C}}({\mathbf{u}},{\mathbf{u}} - {\mathbf{v}}) + {\mathbf{C}}({\mathbf{v}}, {\mathbf{u-v}})],({\mathbf{AB}})^{ - \delta }({\mathbf{u}} - {\mathbf{v}})} \right\rangle _\mathbb{H}  \hfill \\
  {\text{                    }} \leqslant  - \lambda _0^{ - 1} a^{-2\delta}\left\| {{\mathbf{u}} - {\mathbf{v}}} \right\|_\mathbb{H}^2  + ca^{-\delta}{\nu}^{ - 1} \lambda _1^{ - (1+\delta)}\left\| {{\mathbf{u}} - {\mathbf{v}}} \right\|_\mathbb{H}^2 \left( {\left\| {\mathbf{u}} \right\|_\mathbb{H}  + \left\| {\mathbf{v}} \right\|_\mathbb{H} } \right) \hfill \\
  {\text{                    }} \le - \lambda _0^{ - 1} a^{-2\delta} \left\| {{\mathbf{u}} - {\mathbf{v}}} \right\|_\mathbb{H}^2  + ca^{-\delta}{\nu}^{ - 1} \lambda _1^{ - (1+\delta)} \left\| {{\mathbf{u}} - {\mathbf{v}}} \right\|_\mathbb{H}^2 {\mathbf{u}}_ +   \hfill \\
  {\text{                    }} =  - \lambda _0^{ - 1} a^{-2\delta} \left\| {{\mathbf{u}} - {\mathbf{v}}} \right\|_\mathbb{H}^2  + ca^{-\delta}{\nu}^{ - 1} \lambda _1^{ - (1+\delta)} \left\| {{\mathbf{u}} - {\mathbf{v}}} \right\|_\mathbb{H}^2 \left( {\tfrac{1}
{2}\nu \lambda _1^{(1+\delta)}(c^{-1} a^{-\delta}\lambda _0^{ - 1}) \left\{ {1+ \sqrt {1 - \gamma } } \right\}} \right) \hfill \\
  {\text{                    }} =  - \tfrac{1}
{2}\lambda _0^{ - 1} a^{-2\delta}\left\| {{\mathbf{u}} - {\mathbf{v}}} \right\|_\mathbb{H}^2 \left\{ {1 - \sqrt {1 - \gamma } } \right\} \hfill \\
  {\text{                    }} =  - \alpha \left\| {{\mathbf{u}} - {\mathbf{v}}} \right\|_\mathbb{H}^2 ,{\text{   }} \alpha = \tfrac{1}
{2}\lambda _0^{ - 1} a^{-2\delta} \left\{ {1 - \sqrt {1 - \gamma } } \right\}. \hfill \\ 
\end{gathered} 
\eeqa
\end{proof} 
\begin{thm} The operator ${\mathcal{A}}(t) = \nu {\mathbf{A}^{(1+\delta)}} \mathbf{J}( \cdot ,t)$
 is closed, uniformly dissipative and jointly continuous in ${\mathbf{u}}$ and $t$.  Furthermore, for each $t \in {\mathbf{R}}^ +  $ and $\beta  > 0$, 
$Ran[I - \beta  {\mathcal{A}}(t)] \supset \mathbb{B}[ \Om ]$, so that $
 {\mathcal{A}}(t)$ is m-dissipative on $\mathbb{D}$. 
\end{thm}
\begin{proof} 
  Since ${\mathbf{J}}( \cdot ,t)$ is strongly dissipative and closed on $
\mathbb{B}$, it follows from Theorem 6 that $
Ran[{\mathbf{J}}( \cdot ,t)] \supset \mathbb{B}$.

To show that ${\mathcal{A}}(t) = \nu ({\mathbf{AB}})^{(1 + \delta )} {\mathbf{J}}( \cdot ,t)$ is uniformly dissipative for $
{\mathbf{u}},{\mathbf{v}} \in \mathbb{B}{_+}$, we have
\beqa
\left\langle {{\mathcal{A}}(t){\mathbf{u}} - {\mathcal{A}}(t) {\mathbf{v}} , ({\mathbf{u}} - {\mathbf{v}})} \right\rangle _\mathbb{H} 
=  - \nu \left\| {{\mathbf{A}}^{1/2} ({\mathbf{u}} - {\mathbf{v}})} \right\|_\mathbb{H}^2 \qquad \qquad \qquad \qquad \qquad \qquad \hfill \\
 - \left\langle (1/2){[{\mathbf{C}}({\mathbf{u}} - {\mathbf{v}},{\mathbf{u}}) + {\mathbf{C}}({\mathbf{u}} -{\mathbf{v}}, {\mathbf{v}})],({\mathbf{u}} - {\mathbf{v}})} \right\rangle _\mathbb{H}. 
\eeqa
Now, from equation (4), 
\beqa
\left| {\left\langle {[{\mathbf{C}}({\mathbf{u}} - {\mathbf{v}},{\mathbf{u}}) + {\mathbf{C}}({\mathbf{u}} - {\mathbf{v}},{\mathbf{v}})],({\mathbf{u}} - {\mathbf{v}})} \right\rangle _\mathbb{H} } \right| \qquad \qquad \qquad \qquad \qquad \qquad \hfill \\
\leqslant c\left\| {{\mathbf{A}}^{1/2} ({\mathbf{u}} - {\mathbf{v}})} \right\|_\mathbb{H} \left\| {({\mathbf{u}} - {\mathbf{v}})} \right\|_\mathbb{H} \left\{ {\left\| {{\mathbf{Au}}} \right\|_\mathbb{H}  + \left\| {{\mathbf{Av}}} \right\|_\mathbb{H} } \right\}.
\eeqa
We now use $ - \lambda _0^{-1}a^{ - \delta }\left\| {({\mathbf{u}} - {\mathbf{v}})} \right\|_\mathbb{H}  \geqslant  -  \left\| {{\mathbf{A}}^{1/2} ({\mathbf{u}} - {\mathbf{v}})} \right\|_\mathbb{H}$, and the fact that the first eigenvalue of $\bf{B}$ is $1/2$, so that $\lambda_1^{1+\delta}<1$, to get:
\beqa
\begin{gathered}
  \left\langle {{\mathcal{A}}(t){\mathbf{u}} - {\mathcal{A}}(t){\mathbf{v}},{\mathbf{u}} - {\mathbf{v}}} \right\rangle _\mathbb{H}  \leqslant  - \nu \left\| {{\mathbf{A}}^{1/2} ({\mathbf{u}} - {\mathbf{v}})} \right\|_\mathbb{H}^2  + \tfrac{1}
{2}c\left\| {{\mathbf{A}}^{1/2} ({\mathbf{u}} - {\mathbf{v}})} \right\|_\mathbb{H} \left\| {({\mathbf{u}} - {\mathbf{v}})} \right\|_\mathbb{H} \left\{ {\left\| {{\mathbf{Au}}} \right\|_\mathbb{H}  + \left\| {{\mathbf{Av}}} \right\|_\mathbb{H} } \right\} \hfill \\
  {\text{                                   }} = \left\| {{\mathbf{A}}^{1/2} ({\mathbf{u}} - {\mathbf{v}})} \right\|_\mathbb{H} \left\{ { - \nu \left\| {{\mathbf{A}}^{1/2} ({\mathbf{u}} - {\mathbf{v}})} \right\|_\mathbb{H}  + \tfrac{1}
{2}c\left\| {{\mathbf{u}} - {\mathbf{v}}} \right\|_\mathbb{H} \left[ {\left\| {{\mathbf{Au}}} \right\|_\mathbb{H}  + \left\| {{\mathbf{Av}}} \right\|_\mathbb{H} } \right]} \right\} \hfill \\
  {\text{                                    }} \leqslant \left\| {{\mathbf{A}}^{1/2} ({\mathbf{u}} - {\mathbf{v}})} \right\|_\mathbb{H} \left\| {{\mathbf{u}} - {\mathbf{v}}} \right\|_\mathbb{H} \left\{ { - \nu \lambda _0^{-1}a^{ - \delta }  + c{\mathbf{u}_+} } \right\} \hfill \\
    {\text{                                    }} \leqslant \left\| {{\mathbf{A}}^{1/2} ({\mathbf{u}} - {\mathbf{v}})} \right\|_\mathbb{H} \left\| {{\mathbf{u}} - {\mathbf{v}}} \right\|_\mathbb{H} \left\{ {  - \nu \lambda _0^{-1} a^{ - \delta }  + \tfrac{1}
{2}\nu \lambda _1^{ (1+ \delta) } \lambda _0^{-1} a^{ - \delta }\left[ {1 + \sqrt {1 - \gamma } } \right]} \right\} \hfill \\
  {\text{                                    }} < \tfrac{1}
{2}\nu \lambda _0^{-1}a^{ - \delta } \left\| {{\mathbf{A}}^{1/2} ({\mathbf{u}} - {\mathbf{v}})} \right\|_\mathbb{H} \left\| {{\mathbf{u}} - {\mathbf{v}}} \right\|_\mathbb{H} \left\{ { - 1 + \sqrt {1 - \gamma } } \right\} < 0. \hfill \\ 
\end{gathered} 
\eeqa
If we set $
a\left( {\left\| {({\mathbf{u}} - {\mathbf{v}})} \right\|_{\mathbb{H}} } \right) =  - \tfrac{1}
{2}\nu \lambda _0^{-1}a^{ - \delta } \left[ { - 1 + \sqrt {1 - \gamma } } \right]\left\| {{\mathbf{A}}^{1/2} ({\mathbf{u}} - {\mathbf{v}})} \right\|_{\mathbb{H}}$, we have that:
\[
\left\langle {{\mathcal{A}}(t){\mathbf{u}} - {\mathcal{A}}(t){\mathbf{v}},{\mathbf{u}} - {\mathbf{v}}} \right\rangle _{\mathbb{H}}  \leqslant  - a\left( {\left\| {({\mathbf{u}} - {\mathbf{v}})} \right\|_{\mathbb{H}} } \right)\left\| {({\mathbf{u}} - {\mathbf{v}})} \right\|_{\mathbb{H}} .
\] 
It follows that $ {\mathcal{A}}(t)$ is uniformly dissipative.  Since $ - {\mathbf{A}^{(1+\delta)}}$ is m-dissipative, for $\beta  > 0$, $Ran(I + \beta ({\mathbf{AB})^{(1+\delta)}}) = \mathbb{H}$.  As ${\mathbf{J}}$ is strongly dissipative (in the ball of radius $\tfrac{1}
{2}{\mathbf{u}}_+$) and closed, with $Ran[{\mathbf{J}}] \supset \mathbb{B}$, and
${\mathbf{J}}( \cdot ,t):\mathbb{D}\xrightarrow{{onto}}\mathbb{D}$, 
$ {\mathcal{A}}(t)$ is maximal dissipative (in the ball of radius $\tfrac{1}
{2}{\mathbf{u}}_+$), and also closed, so that 
$Ran[I - \beta  {\mathcal{A}}(t)] \supset \mathbb{B}$.  It follows that $ {\mathcal{A}}(t)$ is m-dissipative on $\mathbb{B}$ for each $t \in {\mathbf{R}}^ +  $ (since $\mathbb{H}$ is a Hilbert space).
To see that $ {\mathcal{A}}(t){\mathbf{u}}$ is continuous in both variables, let ${\mathbf{u}}_n ,{\mathbf{u}} \in \mathbb{B}_{\text{ + }} $, $\left\| {\mathbf{A}({\mathbf{u}}_n  - {\mathbf{u}})} \right\|_{\mathbb{H}}  \to 0$
, with $t_n ,t \in I$ and $t_n  \to t$.  Then (see equation (6)) 
\beqa
\begin{gathered}
  \left\| { {\mathcal{A}}(t_n ){\mathbf{u}}_n  -  {\mathcal{A}}(t){\mathbf{u}}} \right\|_\mathbb{H}  \leqslant \left\| { {\mathcal{A}}(t_n ){\mathbf{u}} -  {\mathcal{A}}(t){\mathbf{u}}} \right\|_\mathbb{H}  + \left\| { {\mathcal{A}}(t_n ){\mathbf{u}}_n  -  {\mathcal{A}}(t_n ){\mathbf{u}}} \right\|_\mathbb{H}  \hfill \\
   = \left\| {{\text{[}}\mathbb{P}{\mathbf{f}}(t_n ) - \mathbb{P}{\mathbf{f}}(t)]} \right\|_\mathbb{H}  + \left\| {\nu {\mathbf{A}}({\mathbf{u}}_n  - {\mathbf{u}}) + [{\mathbf{C}}({\mathbf{u}}_n  - {\mathbf{u}},{\mathbf{u}}_n ) + {\mathbf{C}}( {\mathbf{u}},{\mathbf{u}}_n  -{\mathbf{u}})]} \right\|_\mathbb{H}  \hfill \\
   \leqslant d\left| {t_n  - t} \right|^\theta   + \nu \left\| {{\mathbf{A}}({\mathbf{u}}_n - {\mathbf{u}})} \right\|_\mathbb{H}  + \left\| {{\mathbf{C}}({\mathbf{u}}_n  - {\mathbf{u}},{\mathbf{u}}_n ) + {\mathbf{C}}( {\mathbf{u}},{\mathbf{u}}_n  - {\mathbf{u}})} \right\|_\mathbb{H}  \hfill \\
   \leqslant d\left| {t_n  - t} \right|^\theta   + \nu \left\| {{\mathbf{A}}({\mathbf{u}}_n  - {\mathbf{u}})} \right\|_\mathbb{H}  + c \left\| {{\mathbf{A}}({\mathbf{u}}_n  - {\mathbf{u}})} \right\|_\mathbb{H} \left\{ {\left\| {{\mathbf{Au}}_n } \right\|_\mathbb{H}  + \left\| {{\mathbf{Au}}} \right\|_\mathbb{H} } \right\} \hfill \\
   \leqslant d\left| {t_n  - t} \right|^\theta   + \nu \left\| {{\mathbf{A}}({\mathbf{u}}_n  - {\mathbf{u}})} \right\|_\mathbb{H}  +  + 
2c \left\| {{\mathbf{A}}({\mathbf{u}}_n  - {\mathbf{u}})} \right\|_\mathbb{H} {\mathbf{u}}_ +  . \hfill \\ 
\end{gathered} 
\eeqa
It follows that $ {\mathcal{A}}(t){\mathbf{u}}$ is continuous in both variables.
\end{proof}
Since $\mathbb{B}{_+}$ is the closure of $\mathbb{D} = D({\mathbf{A}}) \cap \mathbb{B}$ equipped with the restriction of the graph norm of ${\mathbf{A}}$ induced on $D({\mathbf{A}})$, it follows that $\mathbb{B}{_+}$ is a closed, bounded, convex set.  We now have: 
\begin{thm} For each $T \in {\mathbf{R}}^ +$, $t \in (0,T)$ and ${\mathbf{u}}_0  \in \mathbb{D} \subset \mathbb{B}$, the global in time Navier-Stokes initial-value problem in $\mathbb{R}^3 :$
\beqn
\begin{gathered}
  \partial _t {\mathbf{u}} + ({\mathbf{u}} \cdot \nabla ){\mathbf{u}} - \nu \Delta {\mathbf{u}} + \nabla p = {\mathbf{f}}(t){\text{ in (}}0,T) \times \mathbb{R}^3  , \hfill \\
  {\text{                              }}\nabla  \cdot {\mathbf{u}} = 0{\text{ in (}}0,T) \times \mathbb{R}^3  , \hfill \\
  {\text{                              }}\mathop {\lim }\limits_{\left\| {\mathbf{x}} \right\| \to \infty }{\mathbf{u}}(t,{\mathbf{x}}) = {\mathbf{0}}{\text{ on (}}0,T) \times \mathbb{R}^3  , \hfill \\
  {\text{                              }}{\mathbf{u}}(0,{\mathbf{x}}) = {\mathbf{u}}_0 ({\mathbf{x}}){\text{ in }}\mathbb{R}^3,  \hfill \\ 
\end{gathered} 
\eeqn
 has a unique strong solution ${\mathbf{u}}(t,{\mathbf{x}})$, which is in
 ${L_{\text{loc}}^2}[[0,\infty); {\mathbb {H}^2}]$ and in
$L_{\text{loc}}^\infty[[0,\infty); {\mathbb V}]
\cap \mathbb{C}^1[(0,\infty);{\mathbb H}]$.
\end{thm}
\begin{proof}
Theorem 7 allows us to conclude that, when ${\mathbf{u}}_0  \in \mathbb{D}$, the initial value problem is solved and the solution ${\mathbf{u}}(t,{\mathbf{x}})$ is in $\mathbb{C}^1[(0,\infty);{\mathbb D}]$.  Since $\mathbb{D} \subset \mathbb{H}^{2}$, it follows that ${\mathbf{u}}(t,{\mathbf{x}})$ is also in $\mathbb{ V}$, for each $t>0$.  It is now clear that, for any $T>0$,
\[
\int_0^T {\left\| {{\mathbf{u}}(t,{\mathbf{x}})} \right\|_{\mathbb{H}}^2 dt}  < \infty ,{\text{ and }}\sup _{0 < t < T} \left\| {{\mathbf{u}}(t,{\mathbf{x}})} \right\|_{\mathbb{V}}^2  < \infty .
\]
This gives our conclusion.
\end{proof}
\section*{Discussion}
 It is known that, if ${\mathbf{u}_0} \in \mathbb{V}$, and $\mathbf{f}(t)$ is $L^{\infty}[(0,\infty), \mathbb{H}]$ then there is a time $T> 0$ such that a weak solution with this data is uniquely determined on any subinterval of $[0,T)$ (see Sell and You, page 396, \cite{SY}).   Thus, we also have that: 
\begin{cor} For each $t \in {\mathbf{R}}^ + $ and $
{\mathbf{u}}_0  \in \mathbb{D} $ the Navier-Stokes initial-value problem on $ \mathbb{R}^3 :$
\beqn
\begin{gathered}
  \partial _t {\mathbf{u}} + ({\mathbf{u}} \cdot \nabla ){\mathbf{u}} - \nu \Delta {\mathbf{u}} + \nabla p = {\mathbf{f}}(t){\text{ in (}}0,T) \times \mathbb{R}^3  , \hfill \\
  {\text{                              }}\nabla  \cdot {\mathbf{u}} = 0{\text{ in (}}0,T) \times \mathbb{R}^3  , \hfill \\
  {\text{                              }}\mathop {\lim }\limits_{\left\| {\mathbf{x}} \right\| \to \infty }{\mathbf{u}}(t,{\mathbf{x}}) = {\mathbf{0}}{\text{ on (}}0,T) \times \mathbb{R}^3  , \hfill \\
  {\text{                              }}{\mathbf{u}}(0,{\mathbf{x}}) = {\mathbf{u}}_0 ({\mathbf{x}}){\text{ in }}\mathbb{R}^3  . \hfill \\ 
\end{gathered} 
\eeqn
 has a unique weak solution
${\mathbf{u}}(t,{\mathbf{x}})$, which is in
 ${L_{\text{loc}}^2}[[0,\infty); {\mathbb {H}^2}]$ and in
$L_{\text{loc}}^\infty[[0,\infty); {\mathbb V}]
\cap \mathbb{C}^1[(0,\infty);{\mathbb H}]$.
\end{cor}
Since we require that our initial data be in $\mathbb{H}^{2}$, the conditions for the Leray-Hopf weak solutions are not satisfied.  However, it was an open question as to whether these solutions developed singularities, even if ${\mathbf{u}}_0 \in \mathbb{C}_0^\iy $ (see Giga \cite{G} and references therein).  The above Corollary shows that it suffices that 
${\mathbf{u}}_0 ({\mathbf{x}})\in {\mathbb {H}^2}$ to insure that the solutions develop no singularities.

\acknowledgements
We would like to thank Professor George Sell for his  constructive remarks on an earlier draft.  We have benefited from his friendship, encouragement and the generous sharing of his knowledge over the last fifteen years.  We would also like to sincerely thank Professor Edriss Titi for comments which helped us improve our presentation and eliminate a few areas of possible confusion. 

\end{document}